\documentclass[showpacs,preprint,aps,prl]{revtex4}

\newcommand{\BE}{\begin{equation}}
\newcommand{\EE}{\end{equation}}
\newcommand{\BA}{\begin{eqnarray}}
\newcommand{\EA}{\end{eqnarray}}
\usepackage{graphicx}
\usepackage{dcolumn}
\usepackage{bm}
\input epsf.tex

\begin{document}

\title{Fluctuation and localization of acoustic waves in bubbly water}

\author{Chao-Hsien Kuo}
\author{Ken Kang-Hsin Wang\footnote{Address after August 2003: Department of Physics, University of Rochester, Rochester, NY 14627,
USA}}\author{Zhen Ye}\email{zhen@phy.ncu.edu.tw} \affiliation{Wave
Phenomena Laboratory and Center of Complex Systems, Department of
Physics, National Central University, Chungli, Taiwan 32054,
Republic of China}

\date{July 1, 2003}

\begin{abstract}

Here the fluctuation properties of acoustic localization in bubbly
water is explored. We show that the strong localization can occur
in such a system for a certain frequency range and sufficient
filling fractions of air-bubbles. Two fluctuating quantities are
considered, that is, the fluctuation of transmission and the
fluctuation of the phase of acoustic wave fields. When
localization occurs, these fluctuations tend to vanish, a feature
able to uniquely identify the phenomenon of wave localization.

\end{abstract}

\maketitle

Among many unsolved problems in condensed matter
physics\cite{Rama}, the phenomenon of Anderson localization is
perhaps most intriguing. The concept of localization was first
proposed about 50 years ago for possible disorder-induced
metal-insulator transition\cite{Anderson}. That is, the electronic
mobility in a conducting solid could completely come to a halt
when the solid is implanted with a sufficient amount of randomness
or impurities, in the absence of the electron-electron
interaction. As a result, the electrons tend to stay around the
initial site of release, i.~e. localized, and the envelop of their
wave functions decays exponentially from this site \cite{Lee}. The
physical mechanism behind is suggested to be the multiple
scattering of electronic waves by the disorders.

Ever since the inception of this concept, considerable efforts
have been devoted to observing localization or phenomena possibly
related to localization, as reviewed in, for instance,
Ref.~\cite{Lee,Thouless,loc}. By analogy with electronic waves,
the concept of localization has also been extended to classical
systems ranging from acoustic and electromagnetic waves to
seismological waves in randomly scattering media, yielding a vast
body of literature (Refer to the reviews and textbooks
Refs.~\cite{John,Sheng,Lagen,Seis}). In spite of these efforts,
the phenomenon of localization has remained as a conceptual
conjecture rather than a reality that has been observed
conclusively.

The difficulty in observing electronic localization may be
obvious. In actual measurements, the Coulomb interaction between
electrons is hard to be excluded. With this in mind, it has been
suggested that localization might be easier to observe for
classical waves as wave-wave interactions are often negligible in
these systems. However, classical wave localization may yet suffer
from effects of absorption, leading to uncertainties in data
interpretation. Indeed, up to date, a definite experimental
evidence of localization is still lacking. Significant debates on
reported experimental results remain \cite{McCall,Meade,Wiersma}.
How to unambiguously isolate localization effects from other
effects therefore stays on the task list of top priority, and
poses an compelling issue to our concern.

In this Letter, we wish to continue previous efforts in exploring
acoustic localization in water having many air-filled bubbles,
i.~e. bubbly water \cite{Alberto,APL}. We attempt to identify some
unique features associated with localization. In particular, we
will study the behavior of the phase of the acoustic wave fields
in bubbly water. The fluctuation of the phase and the acoustic
transmission will be studied. The method is based upon the
standard multiple scattering theory which has been detailed in
\cite{Ken}.

Consider the acoustic emission from a bubble cloud in water. For
simplicity, the shape of the cloud is taken as spherical. Such a
model eliminates irrelevant edge effects, and is useful to
separate phenomena pertinent to discussion. Total $N$ bubbles of
the same radius $a$ are randomly distributed within the cloud. The
volume fraction, the space occupied by bubbles per unit volume, is
taken as $\beta$; thus the numerical density of bubbles is $n =
3\beta/(4\pi a^3)$, and the radius of the bubble cloud is $R =
(N/\beta)^{1/3} a$. A monochromatic acoustic source is located at
the center of the cloud. Adaptation of such a model for other
geometries and situations is straightforward. The wave transmitted
from the source propagates through the bubble layer, where
multiple scattering incurs, and then it reaches a receiver located
at some distance from the cloud. The multiple scattering in the
bubbly layer is described by a set of self-consistent equations.
The energy transmission and the acoustic wave field can be solved
numerically in a rigorous fashion \cite{Alberto,Ken}.

Denoting the total acoustic wave at a spatial point as
$p(\vec{r})$, which includes the contributions from the wave
directed from the source ($p_0$) and the scattered waves from all
bubbles. To eliminate the unnecessary geometrical spreading
factor, we normalize the wave field as $T=p/p_0$; therefore $T$ is
dimensionless. The total transmission is defined as $I = |T|^2$.
The average is $\langle I\rangle$ and its coherent portion $I_c
=\langle |T|\rangle^2$; here $\langle\cdot\rangle$ denotes the
ensemble average over the random configuration of bubble clouds.

For the acoustic wave field, we write $p = |p|\exp(i\theta)$ with
$i = \sqrt{-1}$ here; the modulus $|p|$ represents the strength,
whereas $\theta_i$ the phase of the secondary source. We assign a
two dimensional unit vector $\vec{u}_i$, hereafter termed phase
vector, to each phase $\theta_i$, and these vectors are
represented on a phase diagram parallel to the $x-y$ plane. That
is, the starting point of each phase vector is positioned at the
center of individual scatterers with an angle with respect to the
positive $x$-axis equal to the phase, $\vec{u}_i =
\cos\theta_i\hat{x} + \sin\theta_i\hat{y}$. Letting the phase of
the initiative emitting source be zero, i.~e. the phase vector of
the source is pointing to the positive $x$-direction, numerical
experiments are carried out to study the behavior of the phases of
the bubbles and the spatial distribution of the acoustic wave
modulus.

A general aspect of localization is as follows. The acoustic
energy flow is conventionally $\vec{J}\sim
\mbox{Re}[p^\star(-i\nabla)p].$ Taking the field as
$|p|e^{i\theta}$, the current becomes $|p|^2\nabla\theta$.
Therefore when $\theta$ is constant while $|p|\neq 0$, the flow
stops and energy will be localized or stored in space. This
general picture is absent from previous considerations.

A set of numerical experiments has been carried out. In the
simulation, we parameterize the relevant physical quantities as
follows. The frequency is scaled by the radius of bubbles, i.~e.
$ka$. In this way, it turns out that all the simulation is
dimensionless when no absorption is included \cite{Alberto}. The
fluctuation of the phase of the acoustic field at the bubble sites
is defined as $\delta\theta^2 = \left(\frac{1}{N} \sum_{i}^N
\theta_i^2 - (\frac{1}{N}\sum_i^N\theta_i)^2\right)$. It is
conceivable that this fluctuation actually reflects the
fluctuation of the acoustic wave fields in general. Similarly, the
fluctuation of the transmission is $\delta I^2 = \langle
I^2\rangle - \langle I\rangle^2$. The volume fraction is taken as
$\beta = 0.001$.

\begin{figure}[hbt]
\vspace{10pt} \epsfxsize=6in\epsffile{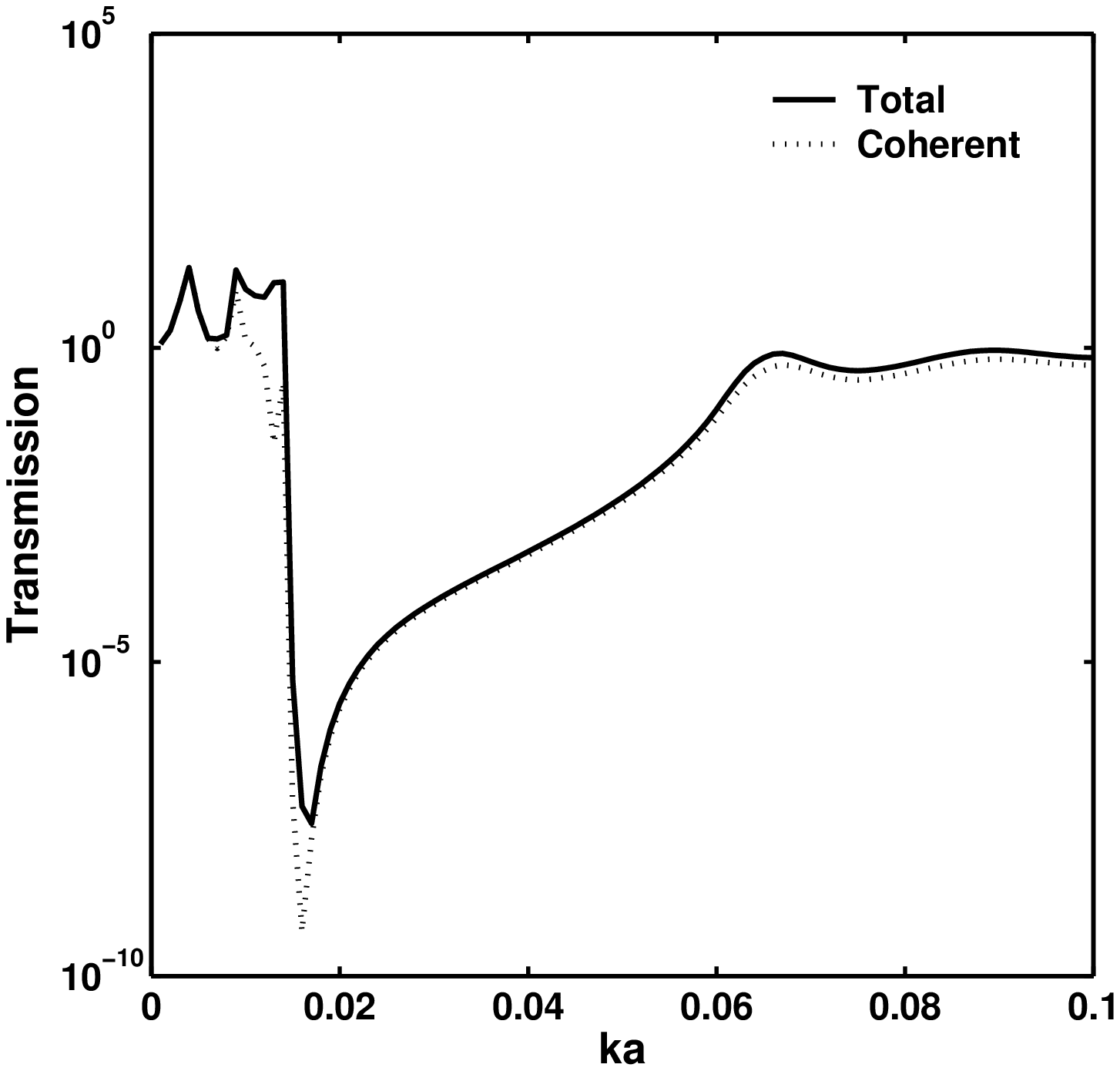}
\caption{Transmissions as a function of $ka$.} \label{fig1}
\end{figure}

First we identify the regimes of localization. Following
\cite{Alberto}, we plot the averaged transmissions ($\langle
I\rangle$ and $I_c$) versus frequency $ka$ in Fig.~\ref{fig1}.
Here $N=800$ and totally 100 averages haven used. The receiver is
located at the distance $2R$ from the source. It is clearly
suggested in the figure that there is a region in which the
transmission is virtually forbidden. This ranges from about $ka =
0.014$ to 0.060. And for most frequencies, the coherent
transmission is the major portion in the transmission.

\begin{figure}[hbt]
\vspace{10pt} \epsfxsize=4in\epsffile{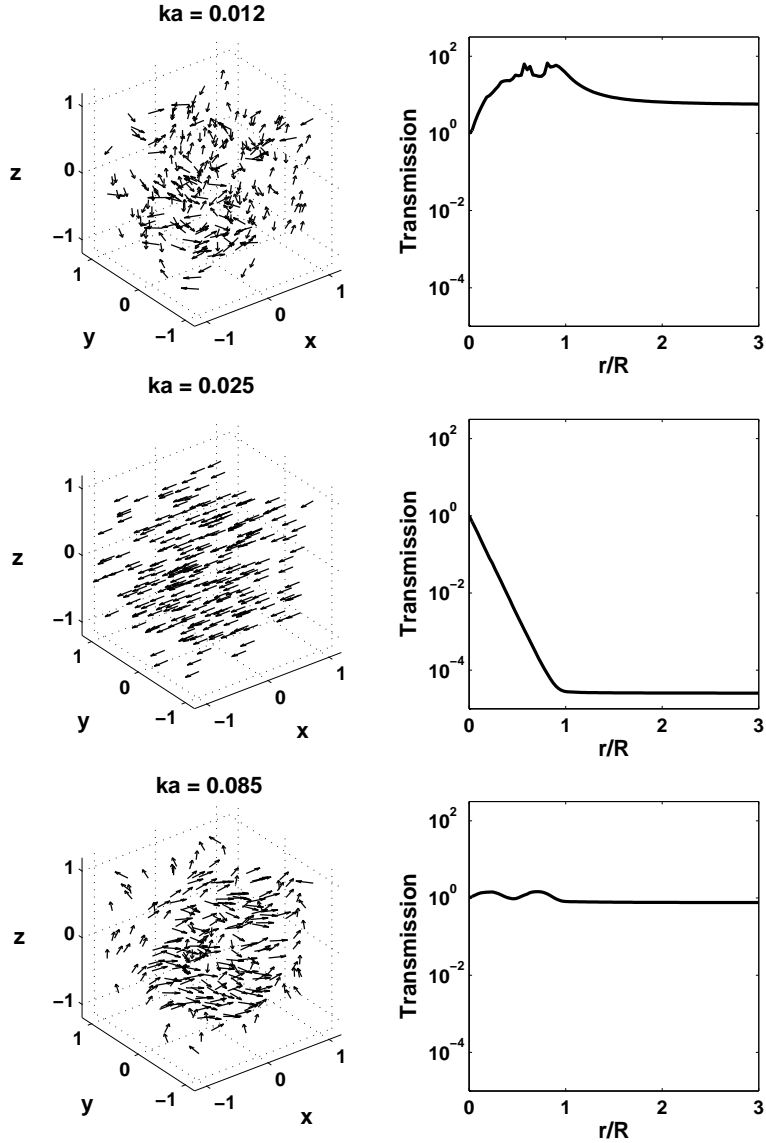} \caption{ Left
column: The phase diagram for the two dimensional phase vectors
defined in the text for an arbitrary random distribution of
bubbles. Right column: The acoustic energy distribution or
transmission, averaged for all directions, as a function of
distance away from the source; the distance is scaled by the
sample size $R$. The three frequencies are selected from within
and outside of the localization regime, referring to
Fig.~\ref{fig1}.} \label{fig2}
\end{figure}

To examine the fluctuation behavior of the phase $\theta$ and the
transmission and for the convenience of the reader, in
Fig.~\ref{fig2}, we purposely repeat the earlier effort\cite{APL}
to plot the phase diagrams of the phase vectors (left column) and
the averaged wave transmission as a function of distance from the
source (right column). Here it is clearly shown that the
exponential decay of the transmission, i.~e. the phenomenon of
localization, is indeed associated with an in-phase coherence or
`ordering' among the phase vectors, i.~e.~nearly all the phase
vectors point to the same direction, in the complete agreement
with the general prescription of localization stated above.

\begin{figure}[hbt] \vspace{10pt}
\epsfxsize=6in\epsffile{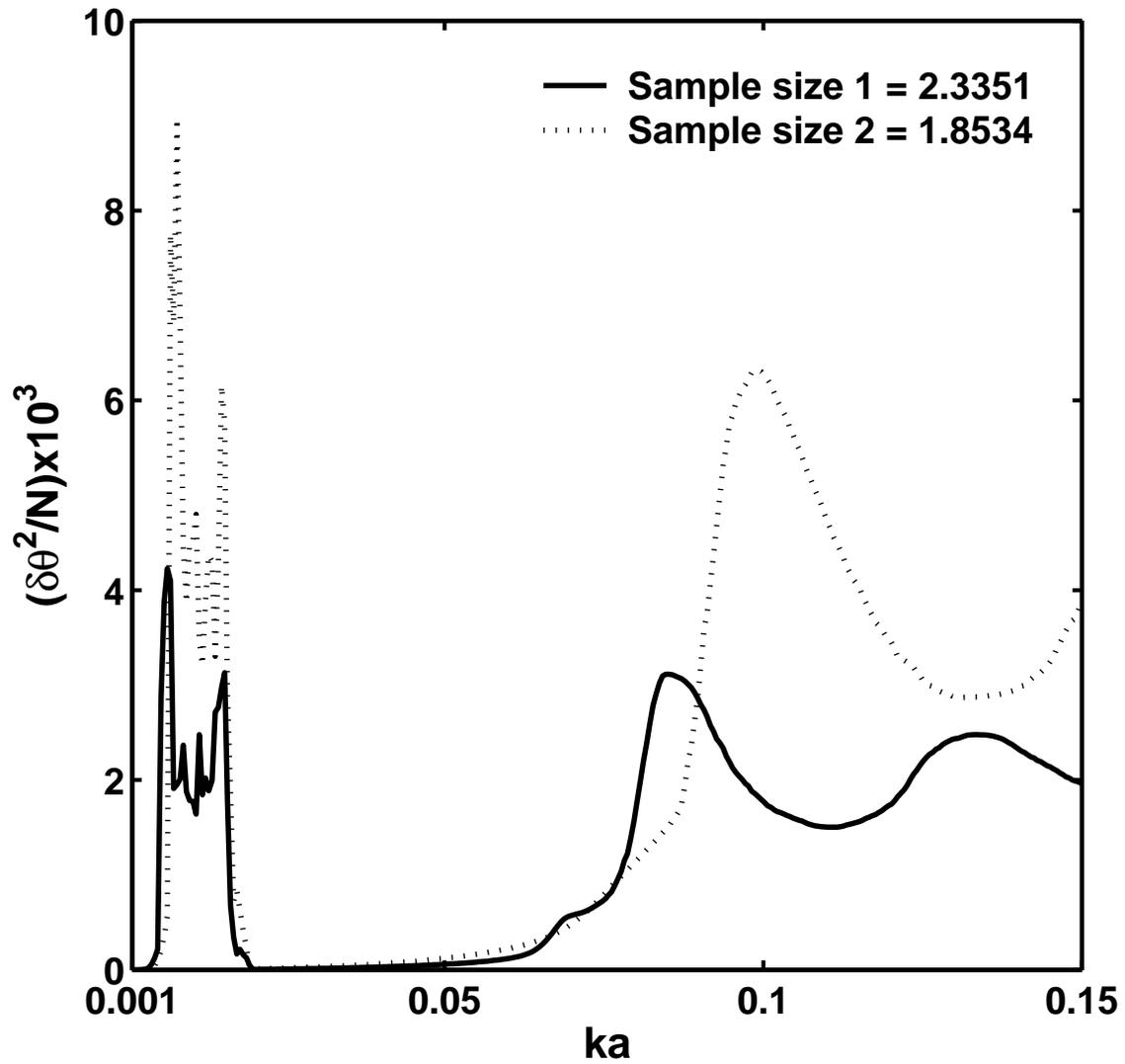} \caption{Fluctuation of the
phase as a function of $ka$ for two sizes of bubble clouds. In
both cases, the volume fraction is kept at 0.001.} \label{fig3}
\end{figure}

In actual experiments such as acoustic scintillation in turbulent
media, it is the variability of signal that is often easier to
analysis. As such, we wish to investigate the fluctuation behavior
of wave transmission. In Fig.~\ref{fig3}, we plot the fluctuation
of the phase of the acoustic wave fields at the bubble sites in
terms of per scatterer. Two sizes of bubble clouds are taken,
corresponding to $N = 800$ and $1600$ respectively. Here we see
that within the localization range, the fluctuation tends to zero
and is insensitive to the sample size. At around the localization
transition edges, significant peaks in the fluctuation appear. The
peak position at the low frequency end does not depend on the
sample size whereas at the high frequency side, the peak position
moves as the sample size increases. These results imply that there
are perhaps two types of transition mechanisms at the low and high
frequency edges respectively.

In the present model we may also be able to include dissipation
effects either manually or through calculation of thermal exchange
and viscosity effects \cite{Ken}. It can be shown that the
dissipation will not be able to give rise to the picture depicted
by Fig.~\ref{fig3}. Therefore these fluctuation behaviors may help
in discerning localization.  We also note here that we have
carried out two types of averaging. One has been defined above.
The second is to look at the phase fluctuation of the wave field
at a fixed spatial point, likely being the case in experiments, we
found that the two approaches nearly produce the same results.

\begin{figure}
\vspace{10pt} \epsfxsize=6in\epsffile{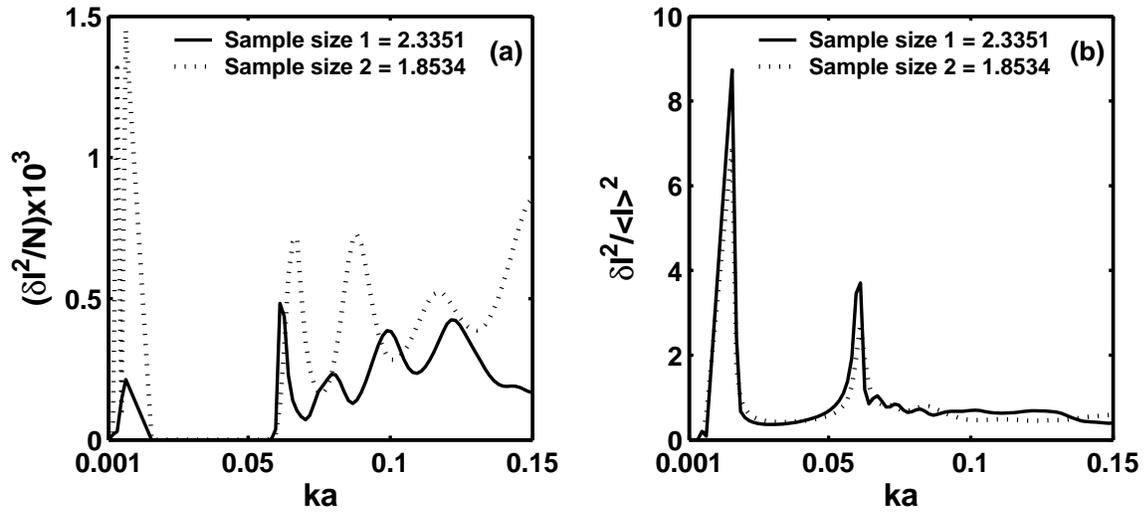} \caption{The
fluctuation of transmission as a function of $ka$.} \label{fig4}
\end{figure}

The corresponding fluctuation of wave transmission per scatterer
$\delta I^2/N$ and the relative fluctuation $\delta I^2/\langle
I\rangle^2$ are plotted against frequency in Fig.~\ref{fig4}.
Again two sample sizes are considered. Fig.~\ref{fig4}(a) shows
that the fluctuation per scatterer behaves similarly as the phase
fluctuation, thereby providing a further check point in discerning
localization effects. However, the relative fluctuation in
Fig.~\ref{fig4}(b) behaves rather differently. There is no
significant reduction within the localization regime, compared to
the outside. Except near the transition edges, the magnitude of
the relative fluctuation is nearly the same for all frequencies
within or outside the localization range and for different sample
sizes, a result in contrast to the discussion for one dimensional
systems \cite{JPC}. We note that at the zero frequency limit, all
the fluctuation approaches zero. This is because at this limit,
the scattering effect is nearly zero, and therefore the bubbles
give no effects.

In short, some fluctuation behaviors in acoustic localization in
bubbly water have been studied. The results suggest that proper
analysis of these behaviors may help discern the phenomenon of
localization in a unique manner.

We thank NSC and NCU for supports.

\newpage

\newpage

\newpage

\newpage

\end{document}